\documentclass[fleqn,10pt]{wlscirep}

\usepackage[utf8]{inputenc}
\usepackage[T1]{fontenc}
\usepackage[hyphenbreaks]{breakurl}

\usepackage{amsmath}
\usepackage{amsfonts}
\usepackage{amssymb}
\usepackage{bbm}
\usepackage{graphicx}
\usepackage{amsthm}
\usepackage{dsfont}
\usepackage{bm}
\usepackage[linesnumbered,algoruled,boxed,lined]{algorithm2e}
\usepackage{multirow}
\usepackage[caption = false]{subfig}
\usepackage{booktabs}
\usepackage{color}
\usepackage{url}

\usepackage{authblk}

\usepackage{array}
\usepackage{epstopdf}
\usepackage{comment}
\usepackage{placeins}

\SetNlSty{}{}{}

\let\oldnl\nl    						
\newcommand\nonl{%
	\renewcommand{\nl}{\let\nl\oldnl}}	

\title{Probabilistic prediction and context tree identification in the Goalkeeper Game}  

\author[1]{Noslen Hern\'{a}ndez}
\author[2,*]{Antonio Galves}
\author[3]{Jesus Garcia}
\author[**]{Marcos Dimas Gubitoso}
\author[4]{Claudia D. Vargas}
\affil[1]{Institut National Polytechnique de Toulouse}
\affil[2]{Universidade de S\~ao Paulo}
\affil[3]{Universidade Estadual de Campinas}
\affil[4]{Universidade Federal do Rio de Janeiro}
\affil[*]{galves@usp.br}
\affil[**]{In memoriam}
\providecommand{\mykeywords}[1]{\textbf{Keywords:} #1}

\begin{abstract}	
	In this article we address two related issues on the learning of probabilistic sequences of events. First, which features make the sequence of events generated by a stochastic chain more difficult to predict. Second, how to model the procedures employed by different learners to identify the structure of sequences of events. Playing the role of a goalkeeper in a video game, participants were told to predict step by step the successive directions -- left, center or right -- to which the penalty kicker would send the ball. The sequence of kicks was driven by a stochastic chain with memory of variable length. Results showed that at least three features play a role in the first issue: 1) the shape of the context tree summarizing the dependencies between present and past directions; 2) the entropy of the stochastic chain used to generate the sequences of events; 3) the existence or not of a deterministic periodic sequence underlying the sequences of events. Moreover, evidence suggests that best learners rely less on their own past choices to identify the structure of the sequences of events. 
	
\end{abstract}

\begin{document}

	\flushbottom
	\maketitle
	%
	%
	\thispagestyle{empty}

\mykeywords{statistical learning, probabilistic sequences, context tree models, probability matching, probability maximizing}

\section*{Introduction}
The aim of this work is to model the performance of a player trying to guess successive choices displayed by an electronic video game called the Goalkeeper Game (\url{https://game.numec.prp.usp.br/}). In this game, the participant, playing the role of a goalkeeper, has to guess at each trial the next direction to where the penalty kicker will send the ball. An animation feedback then shows to which direction the ball was actually sent. The sequence of kicks is selected by a stochastic chain with memory of variable length. 

Stochastic chains with memory of variable length were introduced by Rissanen (1983) \cite{Rissanen:83} as a universal model for data compression. Rissanen observed that very often in experimental datasets composed by sequences of symbols, each new symbol appears to be randomly selected by taking into account a sequence of past units whose length is variable and changes as a function of the sequence of past units itself. Rissanen called a \textit{context} the smallest sequence of past symbols required to generate the next one. The set of contexts can be represented by a rooted and labeled oriented tree, henceforth called \textit{context tree}. The procedure to generate the sequence of symbols is defined by the context tree and an associated family of transition probabilities used to choose each next symbol, given the context associated to the sequence of past symbols at each time step. From now on, stochastic chains with memory of variable length will be called context tree models. Under suitable continuity conditions, stationary stochastic chains can be well approximated by a context tree model \cite{Fernandez02}. For that reason, they have been largely used to model biological and linguistic phenomena \cite{Buhlmann-Wyner:99,csiszar_context_2006,Leonardi06, GarivierLeonardi11,galves:12,Galves-Loch:13,BelloniOliveira17,duarte_retrieving_2019,Hernandez2021}. In the experimental protocol considered here the sequences of directions chosen by the kicker have been generated by context tree models.  

In the Goalkeeper game, playing the role of the goalkeeper, the volunteer was instructed to stop the penalty kicks. Obviously, the intrinsic randomness of the algorithm used by the kicker to choose the directions makes it impossible to stop all the penalty kicks. However, the full identification of the context tree and the associated family of probability distributions used by the kicker is an important asset to increase the goalkeeper's success rate. Moreover, adopting a good strategy to face the randomness of the kicker choices might maximize the goalkeeper success rate.   

Actually, two strategies have been proposed to address the problem of making correct guesses in sequences produced by stochastic chains. The kernel of the problem is to identify the structure of the chain in spite of the intrinsic randomness of its realization (see for instance, Schulzea et al.\cite{Schulze2020} and Koehler et al. \cite{Koehler2010}). The first strategy, known as \textit{maximizing strategy}, corresponds to always choosing the outcome with the higher probability. In the second strategy, called \textit{matching strategy}, the participant tries to emulate the selection procedure used to generate the sequences of events.    

In our experimental protocol an extra difficulty appears, namely the fact that the probability distributions used by the kicker depend on the successive contexts occurring in the sequence of his previous choices. This means that the goalkeeper must deal simultaneously with the problem of identifying the contexts and its associated transition probabilities as well as the problem of choosing a strategy. A double problem of this type was already considered by Wang et al. \cite{wang_learning_2017}. 

In this article we address two related issues. First, which features of the stochastic chain generating the sequences of events make it more difficult to predict. Second, how to model the procedures employed by different learners to identify the structure of sequences of events. This is done through a rigorous statistical procedure to identify both the context tree and the strategy used by the goalkeeper to make his guesses. We collected data from 122 participants, each one playing the role of the goalkeeper against a kicker that used one out of four different context tree models. By analyzing their sequences of responses, we investigate whether they correctly identify the context tree model used by the kicker and which strategy they use to face the randomness of the kicker's choices.
	
\section*{Results}

\begin{figure}[p]
	\includegraphics[width = 0.9\linewidth]{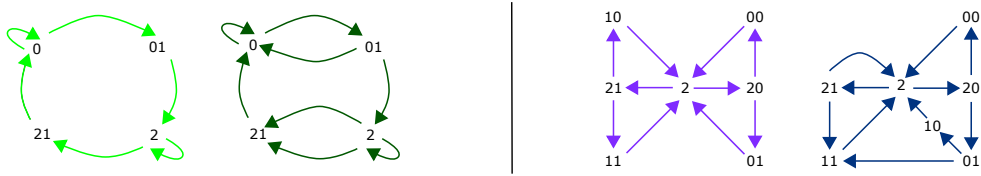}
	\caption{(A) Acting as a goalkeeper, the participant must guess, at each step, to where the next penalty kick will be shot by pressing the corresponding keyboard arrow. The options are left, center or right, represented by the symbols 0, 1, and 2, respectively. An animation feedback shows to which direction the ball was effectively sent. (B) Context tree models governing the kicker’s choices and their corresponding entropy values. (C) Examples of sequences selected by the kicker using each one of the four context tree models. (D) Graph representation of the context tree models governing the kicker’s choices.}	
	\label{fig:exp_design}
\end{figure}

The aim of the experiment was to model the performance of a player trying to guess successive symbols displayed by an electronic video game called the Goalkeeper Game (\url{https://game.numec.prp.usp.br/demo}). Playing the role of a goalkeeper, the participant was told to guess one of the three directions to where the penalty kicker could send the ball: left, center, or right, hereafter represented by the numbers 0, 1, and 2, respectively. An animation feedback showed in which direction the ball was effectively sent (Figure \ref{fig:exp_design}A). 

The sequences of shot directions were generated by four different context tree models (Figure \ref{fig:exp_design}B). Context tree models are characterized by two elements. The first element is a context tree and the second element is a family of transition probabilities indexed by the leaves of the context tree. In our experimental protocol, the four context tree models characterizing the sequences of the kicker's choices will be denoted by $(\tau^k_1, p^k_1),(\tau^k_2, p^k_2),(\tau^k_3, p^k_3), (\tau^k_4, p^k_4)$. The upper index $k$ in the above notation stands for \textit{kicker}. These four context tree models are represented in Figure \ref{fig:exp_design}B. Sequences generated by using each of these context tree models are depicted in Figure \ref{fig:exp_design}C. 

For a fixed context tree and two different associated families of transition probabilities, we conjecture that the context tree model with higher entropy would be more difficult to learn. For the first pair (Figure \ref{fig:exp_design}B, left panel), changes in the transition probabilities associated to the contexts 01 and 21 increased the entropy values from 0.65 in $(\tau^k_1, p^k_1)$ to 0.81 in $(\tau^k_2, p^k_2)$.

We also conjectured that for a fixed context tree and two different associated families of transition probabilities, the one that displays a periodic structure would be easier to learn. For the second pair (Figure \ref{fig:exp_design}B, right panel), sequences generated by the context tree model $(\tau^k_3, p^k_3)$  can be described as a concatenation of strings 211 in which the symbol 1 is replaced by the symbol 0 with a small probability in an i.i.d way. For the context tree model $(\tau^k_4, p^k_4)$, the interchange of transition probabilities associated to 01 and 21 as well as of the most probable outcome of context 2 disrupts the periodic structure displayed in the context tree model $(\tau^k_3, p^k_3)$ without changing the entropy values (0.54 for the context tree model $(\tau^k_3, p^k_3)$ and 0.56 for the context tree model $(\tau^k_4, p^k_4)$). Finally, comparing the performance obtained with the context tree models $(\tau^k_1, p^k_1)$ and $(\tau^k_2, p^k_2)$ with the context tree models $(\tau^k_3, p^k_3)$ and $(\tau^k_4, p^k_4)$ might give an indication that augmenting the number of contexts increases the learning difficulty.

A total of 122 participants was divided into four groups of 30, 31, 31 and 30, respectively. Each context tree model in Figure \ref{fig:exp_design}B was played by a different group of participants (see section Methods). For each participant a sample was constituted by collecting an ordered sequence of 1000 pairs in which the first element at each pair indicates the choice of the kicker at that step and the second element corresponds to that of the goalkeeper.

\subsection*{Time evolution of the performance per context tree model}

\begin{figure}[h!]
	\includegraphics[width = \textwidth]{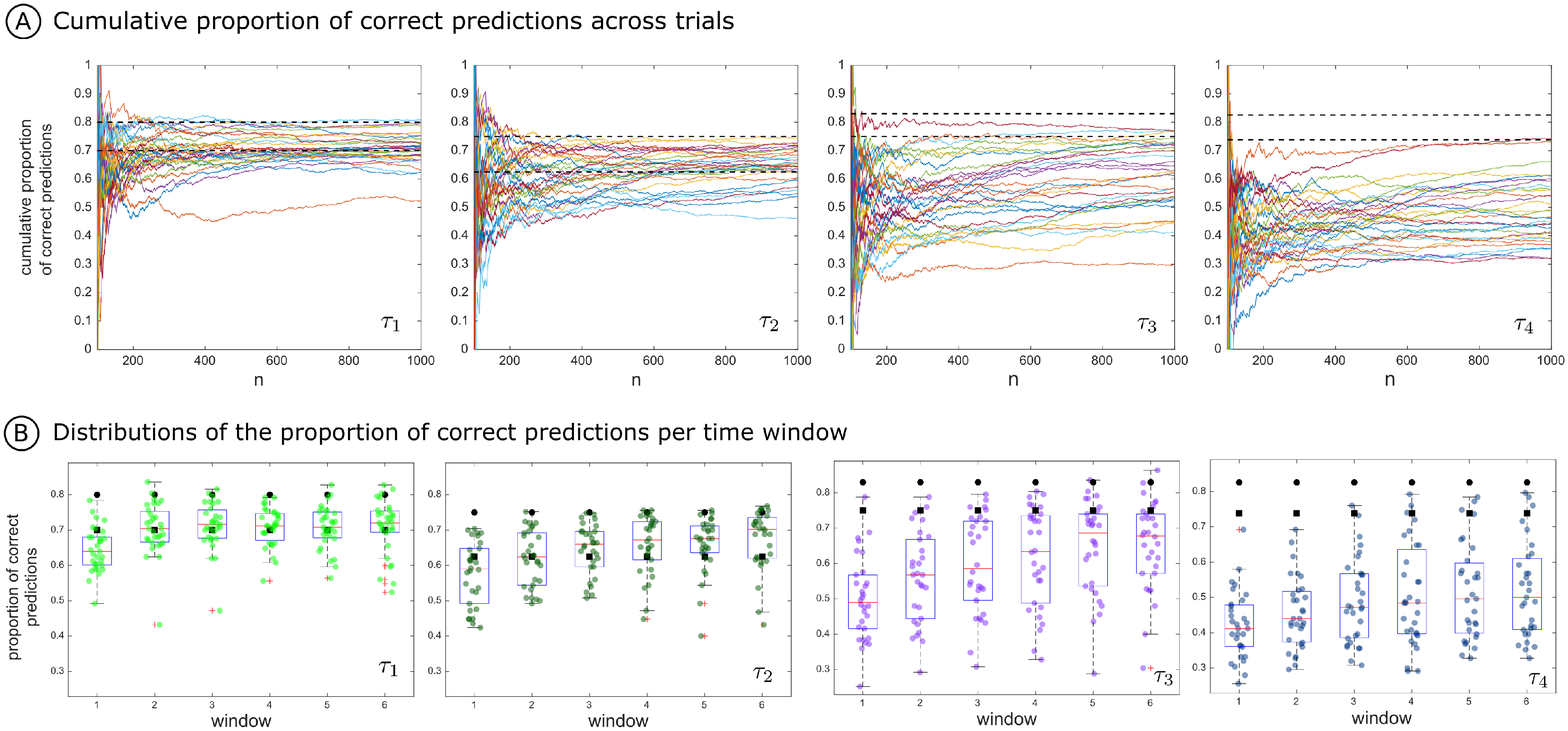}
	\caption{(A) Time evolution from trial 100 to trial 1000 of the cumulative proportion of correct guesses for each context tree model. (B) Boxplots of proportions of correct guesses across participants in a sliding window of length 250 pacing at 150 trials for each context tree model. The proportions of correct guesses that could be achieved by a goalkeeper using the matching (bottom line in (A) and black square marker in (B)) and the maximizing (top line in (A) and black circle in (B)) strategies are indicated.}
	\label{fig:rawdata}
\end{figure}

Figure \ref{fig:rawdata}A shows the cumulative proportion of correct predictions across trials per participant for the four context tree models. An exploratory analysis of the cumulative proportion of correct predictions for models $(\tau^k_1, p^k_1)$ and $(\tau^k_2, p^k_2)$ reveals that the participants tend to lie mostly between the matching and the maximizing strategy scores as the number of trials increases. This is not the case for models $(\tau^k_3, p^k_3)$ and $(\tau^k_4, p^k_4)$. 

A sliding window approach was employed to further explore the temporal evolution of the participants performance for each context tree model. Boxplots (Figure \ref{fig:rawdata}B) depict the distributions of the proportions of correct predictions across participants for each time window and each context tree model. For $(\tau^k_1, p^k_1)$ and $(\tau^k_2, p^k_2)$ the interquartile range is almost above the matching strategy score from the third time window on. 

For $(\tau^k_1, p^k_1)$ and $(\tau^k_2, p^k_2)$ the median of the proportion of correct predictions across participants is above the theoretical matching strategy score from the third time window on. Also, the interquartile range of proportion of correct predictions for $(\tau^k_2, p^k_2)$ is larger than for $(\tau^k_1, p^k_1)$, suggesting a higher performance variability. 

For $(\tau^k_3, p^k_3)$ and $(\tau^k_4, p^k_4)$ the median of the proportion of correct predictions across participants is smaller than the theoretical matching strategy score in all time windows. In $(\tau^k_3, p^k_3)$, the third quartile of the boxplot almost reaches the theoretical matching strategy score from the fourth time window on. Results are even worse for $(\tau^k_4, p^k_4)$ as the third quartile is always clearly below the theoretical matching strategy score. 

Finally, there is a much greater variability in the distribution of proportions of correct predictions across participants in $(\tau^k_3, p^k_3)$ and $(\tau^k_4, p^k_4)$, as compared with $(\tau^k_1, p^k_1)$ and $(\tau^k_2, p^k_2)$. Curiously, for $(\tau^k_1, p^k_1)$ there are some outliers in time window 6, suggesting that the performance of some participants deteriorated towards the end of the task.    

\FloatBarrier

\subsection*{Identifying the goalkeeper strategy}

\begin{figure}[h!]
	\includegraphics[width = \textwidth]{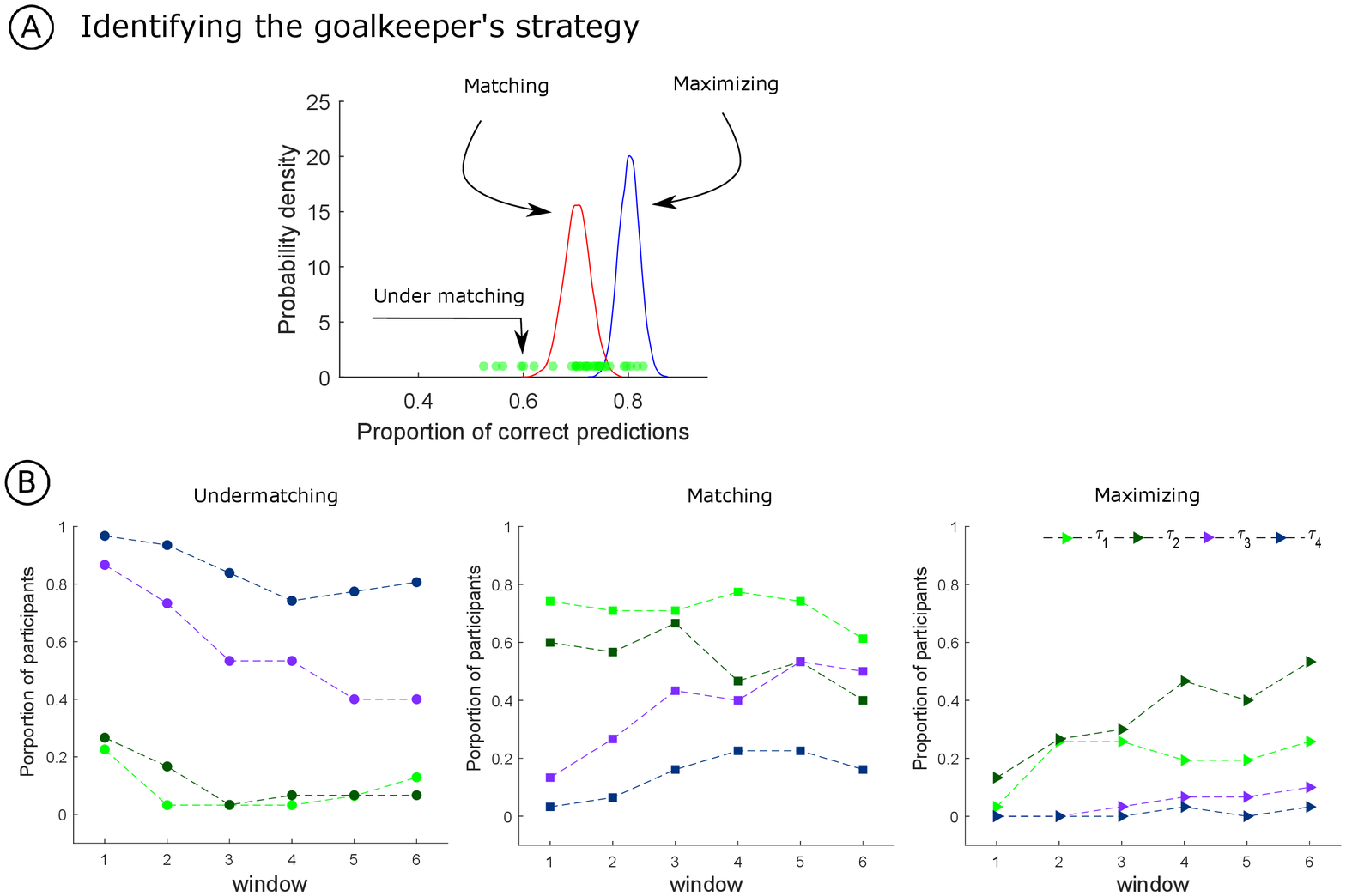}%
	\caption{A) A probability density of the proportion of correct predictions for matching and  maximizing strategies was estimated using a kernel density estimator on simulated data. For each participant and each window of analysis, the likelihood that the participant’s proportion of correct guesses was generated by one of the two estimated distributions (matching, in red vs. maximizing, in blue) is considered to decide to which strategy the participant is close to B) Proportion of participants per window of analysis that undermatched (left) matched (center) and maximized (right) per context tree model.}    
	\label{fig:strategies}%
\end{figure}

To identify the strategy to which a given participant was closer to, we estimated, for each context tree model, a probability density of the proportion of correct predictions for the matching and the maximizing strategies. This was done by comparing, for each participant and each window of analysis, the likelihood that the participant's proportion of correct guesses was generated by one of the two distributions (matching vs. maximizing). See Figure \ref{fig:strategies}A. 

Two samples of proportions of correct predictions corresponding to a goalkeeper using the matching and the maximizing strategies were simulated. This was done by generating 10000 kicker sequences of size 250 (the size of each windows of analysis) and the corresponding response sequences. Then a kernel density estimator was used to obtain a probability density estimate for each strategy. 

Figure \ref{fig:strategies}B depicts the proportion of participants per window of analysis that employed undermatching, matching, and maximizing strategies per context tree model. For $(\tau^k_1, p^k_1)$ and $(\tau^k_2, p^k_2)$ the great majority of participants laid either at the matching or the maximizing strategies in all time windows. Interestingly for $(\tau^k_2, p^k_2)$ the proportion of participants employing the matching considerably reduced in favor of the maximizing strategy across time. For $(\tau^k_3, p^k_3)$ most participants started by employing an undermatching strategy which was succeeded progressively by a matching strategy. Finally for $(\tau^k_4, p^k_4)$ the undermatching strategy prevailed across time. Almost no participant achieved the maximizing strategy for $(\tau^k_3, p^k_3)$ and $(\tau^k_4, p^k_4)$. 

\FloatBarrier

\subsection*{ANOVA of participants' performance across time windows}

To access the differences in performance between the context tree models across time windows, a statistical analysis was done using a two-way mixed ANOVA. The intrinsic randomness of each of the context tree models used to guide the choice of the kicker implies that the optimal performance associated to the maximizing strategy differs from one model to another (see top dashed lines in Figure \ref{fig:rawdata}A). Therefore, for statistical analysis, the proportions of correct predictions obtained per participant and per time window were normalized using the theoretical maximizing strategy score of the corresponding context tree model. These normalized proportions of correct guesses were transformed using a logit transformation (see Supplementary Figure S1).               

To eliminate data outliers, an univariate linear regression model was fitted to each participant's normalized proportions of correct guesses (in logit scale) as a function of the time window. Participants displaying a negative slope in the estimated regression line were excluded from the subsequent analysis (see Supplementary Figure S2). As a consequence, the final number of participants per context tree models used in the analysis are 24, 24, 27 and 26, respectively.    

The two-way mixed ANOVA analysis of the participants' normalized proportions of correct guesses (in logit scale) considers the context tree model as a between subject factor and the time window as a within subject factor. In our case, the levels of the between subject factor were $(\tau^k_1, p^k_1), (\tau^k_2, p^k_2), (\tau^k_3, p^k_3), (\tau^k_4, p^k_4)$ and the levels of the within subject factor are $1, 2, 3, 4, 5, 6$. A significant interaction between the time window and the context tree model, $F(265.67, 8.22) = 3.04, p=0.003$, indicated that the performance evolved differently across the four context tree models.

\begin{figure}[h!]
	\centering
	\includegraphics[width = 0.5\textwidth]{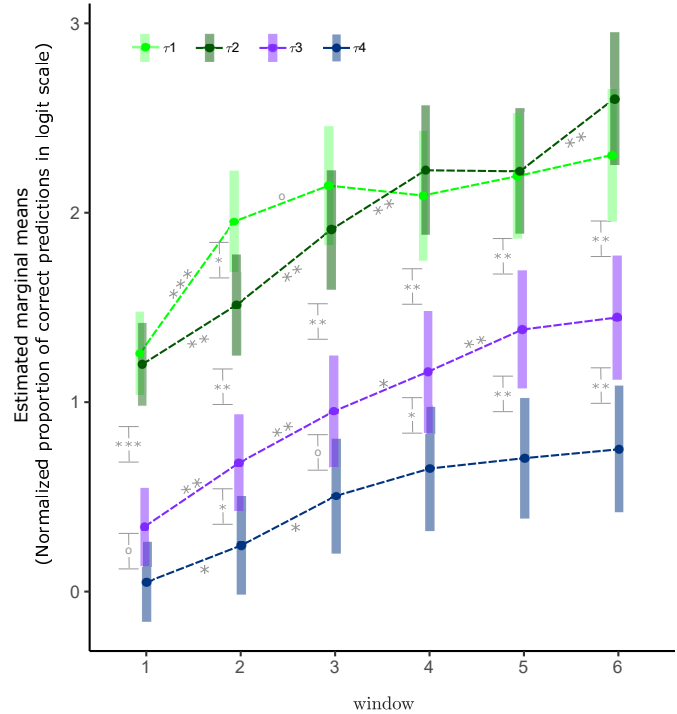}%
	\caption{Interaction graph corresponding to the two-way mixed ANOVA analysis using the logit transformation of the normalized proportions of correct predictions as dependent variable and the context tree and time window as factors. Marginal means and 95\% confidence intervals of the means are represented with dots and bars, respectively. For each context tree model, the significance level of the difference between successive time windows is indicated using the following convention: $***$ for a p-value in the interval $[0,0.0001)$, $**$ for a p-value in the interval $[0.0001,0.01)$, $*$ for a p-value in the interval $[0.01,0.05)$, $\circ$ for a p-value in the interval $[0.05,0.1)$, null for a p-value in the interval $[0.1,1]$. The same convention is used to indicate the significant level of the difference between the means of $(\tau^k_1, p^k_1)$ and $(\tau^k_2, p^k_2)$, $(\tau^k_2, p^k_2)$ and $(\tau^k_3, p^k_3)$, and $(\tau^k_3, p^k_3)$ and $(\tau^k_4, p^k_4)$, for each time window. }    
	\label{fig:anova}%
\end{figure}

Figure \ref{fig:anova} shows the graph of interactions of the two-way mixed ANOVA analysis. The differences between the means at consecutive time windows per context tree model were tested to access the performance evolution for that context tree model. A comparison of the means of the context tree models $(\tau^k_1, p^k_1)$ versus $(\tau^k_2, p^k_2)$, $(\tau^k_2, p^k_2)$ versus $(\tau^k_3, p^k_3)$ and $(\tau^k_3, p^k_3)$ versus $(\tau^k_4, p^k_4)$ was performed per time window. To globally control the level of significance of the test with multiple comparisons, the Benjamini \& Hochberg correction was used \cite{Benjamini1995}. 

\FloatBarrier
 

For model $(\tau^k_1, p^k_1)$, the participants' performance strongly improved from the first to the second time window and then stabilized, with no more significant improvement (see Figure \ref{fig:anova} and Supplementary Table S1 for exact p-values). Conversely, for model $(\tau^k_2, p^k_2)$, significant differences appeared up to the fourth time window, then the performance stabilized and presented a significant improvement in the step to the last time window (see Figure \ref{fig:anova} and Supplementary Table S1 for exact p-values). Besides, comparison of $(\tau^k_1, p^k_1)$ and $(\tau^k_2, p^k_2)$ performance per time window revealed that the only significant difference occurs at the second time window. Changing the transitions associated to contexts 01 and 21 from deterministic in $(\tau^k_1, p^k_1)$ to random in $(\tau^k_2, p^k_2)$  increases the entropy of the corresponding stochastic chains from 0.65 to 0.81. As a consequence, the participants needed more time to learn the structure of the chain.

For model $(\tau^k_3, p^k_3)$, the performance of the participants improved significantly up to the fifth time window. For $(\tau^k_4, p^k_4)$, significant differences were detected only up to the third time window. Besides, $(\tau^k_3, p^k_3)$ significantly differed from $(\tau^k_4, p^k_4)$ in almost all time windows (see Figure \ref{fig:anova} and Supplementary Table S1 for exact p-values). These results suggest that changes made to model $(\tau^k_3, p^k_3)$ to obtain model $(\tau^k_4, p^k_4)$ imposed a significant learning difficulty to model $(\tau^k_4, p^k_4)$ in comparison to model $(\tau^k_3, p^k_3)$. 

Significant differences in performance also appeared between $(\tau^k_2, p^k_2)$ and $(\tau^k_3, p^k_3)$ for all time windows. Thus, differences in performance can be assumed to occur between $\{(\tau^k_1, p^k_1), (\tau^k_2, p^k_2)\}$ and $\{(\tau^k_3, p^k_3), (\tau^k_4, p^k_4)\}$.

\subsection*{Does the goalkeeper identify the context tree used by the kicker?}

To retrieve the structure of the context tree governing the goalkeeper choices from the collected data, we introduce a statistical model selection procedure (see Methods, section Statistical model selection procedure), performed separately for each participant data and each time window. Using this statistical procedure, we retrieved the context tree and the associated family of transition probabilities employed by a goalkeeper $v \in V$,  $(\hat{\tau}^{v,j}_i, \hat{q}^{v,j}_i)$, when the kicker model is $(\tau^k_i, p^k_i), \ i = 1,...,4$ and the time window is $j \in \{1,2,3,4,5,6\}$.

For each context tree model $(\tau^k_i, p_i^k)$ and each time window $j$, we end up with a set of trees $\{\hat{\tau}^{v,j}_i, v \in V_i\}$, where $V_i$ is the subset of participants that played against the kicker using the context tree model $(\tau^k_i, p_i^k)$. The mode context tree \cite{Hernandez2021} of this set of trees is computed to summarize the result of the set of participants.

\begin{figure}[h!]
	\centering
	\includegraphics[width = 0.8\textwidth]{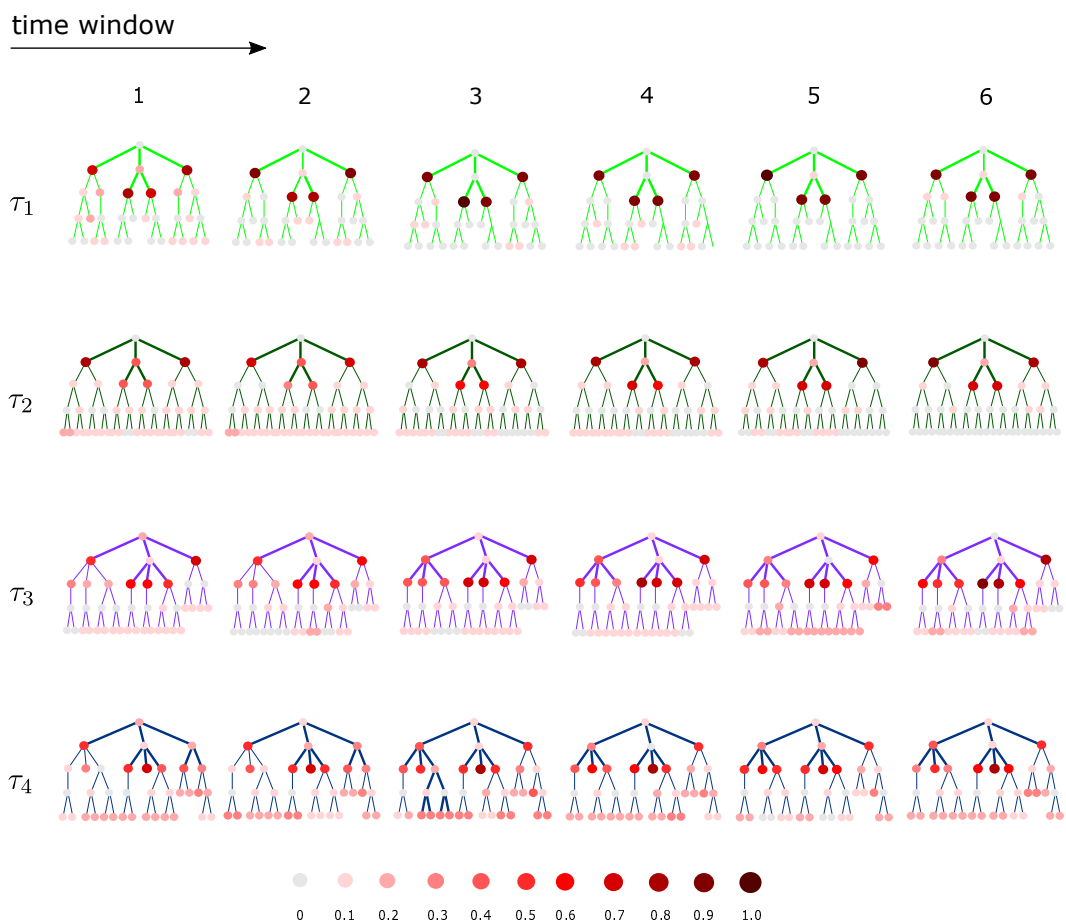}
	\caption{The context trees modeling the goalkeepers choices are summarized for each of the four context tree models. To identify these models we used the responses of each goalkeeper and the kicker choices within a sliding window of length 250 pacing at 150 trials. The nodes at each tree structure represent the strings that different goalkeepers identified as a context. Each node is colored from light pink to dark red according to the proportion of participants identifying the node as a context. Thick lines highlight the mode context tree. The leaves of the mode context tree are the strings that were identified as contexts more often across participants.}
	\label{fig:modecontexttree}
\end{figure}

Figure \ref{fig:modecontexttree} presents the mode context tree computed per time window for each context tree model. This is highlighted in a tree structure that contains all possible past strings up to length 4 that can be identified as a context. 

It can be verified that the mode context tree matches that of the kicker’s context tree as early as in the first time window for models $(\tau_1^k, p_1^k)$ and $(\tau_2^k,p_2^k)$. Nevertheless, a greater consensus around the contexts used by the kicker is observed in $(\tau_1^k, p_1^k)$ than in $(\tau_2^k,p_2^k)$ for all time windows. This suggests that context tree model $(\tau_2^k,p_2^k)$ is more difficult to learn than context tree model $(\tau_1^k, p_1^k)$.  

For models $(\tau_3^k, p_3^k)$ and $(\tau_4^k,p_4^k)$, the mode context tree matches that of the kicker’s context tree in the third and fourth time window, respectively. The fact that a higher number of participants misidentified the kicker's contexts indicates that these models are more difficult to learn. 

\FloatBarrier

\subsection*{Does the goalkeeper rely on theirs own past predictions?}

\begin{figure}[h!]
	\centering
	\includegraphics[width = \textwidth]{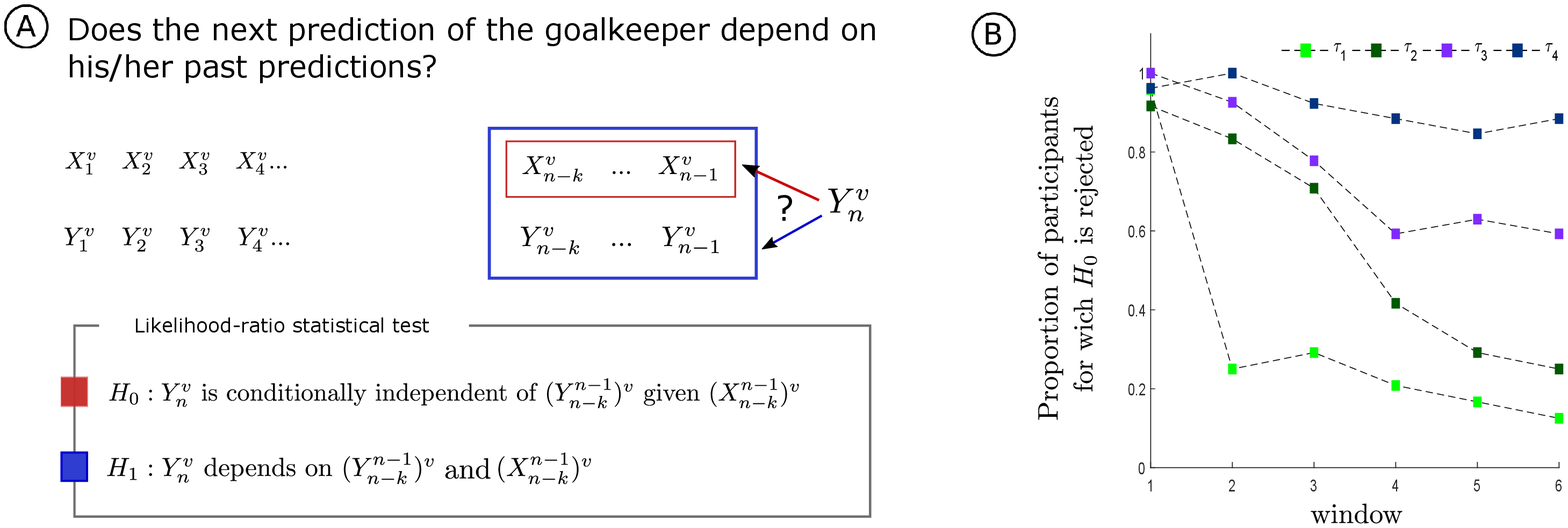}%
	\caption{A) Statistical test to verify  if  the next goalkeeper’s prediction depends or not on his past choices. B) Proportion of participants per context tree model across time windows for which the current choice of the goalkeeper depends both on the past choices of the kicker and his/her past choices.}%
	\label{fig:strategy}%
\end{figure}

The model selection procedure used to retrieve the structures governing the goalkeeper choices assumes that the goalkeeper takes into account the past events of the kicker's sequence. To inspect whether the goalkeeper's predictions are also influenced by their own past choices we performed a statistical hypothesis test. More specifically, we tested the null hypothesis that the goalkeeper's predictions depend only on the past choices of the kicker by means of a likelihood ratio test (see Figure \ref{fig:strategy}A and the Methods, section Likelihood-ratio statistical tests for independence). 

Figure \ref{fig:strategy}B summarizes the results of the hypothesis test. The more the participants have learned the model, the higher is the probability of not rejecting the null hypothesis, indicating that the participants stop being influenced by their own past choices. Our data show that most participants stopped looking at their own past as early as the second window of analysis for the context tree model $(\tau^k_1, p^k_1)$, while for the context tree model $(\tau^k_2, p^k_2)$ this happened only from the fourth window of analysis on. Moreover, for the context tree models $(\tau^k_3, p^k_3)$ and $(\tau^k_4, p^k_4)$, most goalkeepers kept looking at past predictions throughout the game, while trying to identify the tree used by the kicker. This was most evident for the context tree model $(\tau^k_4, p^k_4)$. Taken together, these results suggest that the goalkeeper’s decision seems influenced by theirs own past choices while he/she is still learning the context tree model structure.

\FloatBarrier

\section*{Discussion}


We employed a new mathematical framework to model the relationship between a sequence of events generated by four different context tree models and the goalkeeper's responses. A model selection procedure allowed to retrieve the context tree and the family of transition probabilities governing the predictions of the goalkeeper. 

We were interested in testing whether the goalkeeper performance was affected by the entropy of the kicker context tree models. For a fixed context tree and two different associated families of transition probabilities, the context tree model with higher entropy ($(\tau^k_2, p^k_2)$) was expected to be more difficult to learn. Statistical analysis revealed that participants reached a learning plateau for $(\tau^k_1, p^k_1)$ earlier than for $(\tau^k_2, p^k_2)$. Besides, most participants stopped looking at their own past as early as the second window of analysis for the context tree model $(\tau^k_1, p^k_1)$ while for the context tree model $(\tau^k_2, p^k_2)$, this happened only from the fourth window of analysis on. Taken together, these results indicate that for a fixed context tree, a greater entropy induced by a modification of the associated transition probabilities slows down learning.   

For a fixed context tree and two different associated families of transition probabilities, namely $(\tau^k_3, p^k_3)$ and $(\tau^k_4, p^k_4)$, the one that does not display a periodic structure ($(\tau^k_4, p^k_4)$) was expected to be more difficult to learn comparatively with the one which produces sequences that can be seen as a concatenation of the same deterministic sequence with three elements 211, in which the symbol 1 can be replaced by a new symbol 0 in an i.i.d. way. Indeed, learning rates were lower for the context tree model $(\tau^k_4, p^k_4)$ as compared to the context tree model $(\tau^k_3, p^k_3)$. Furthermore, the mode context tree of the goalkeeper matched that of the kicker from the third window of analysis on for the context tree model $(\tau^k_3, p^k_3)$, and from the fourth window of analysis on for the context tree model $(\tau^k_4, p^k_4)$. For both context tree models, most goalkeepers kept looking at past predictions throughout the game while trying to identify the tree used by the kicker. This was most evident for the context tree model $(\tau^k_4, p^k_4)$.

Finally, the comparison of the performances obtained with the context tree models $(\tau^k_1, p^k_1)$ and $(\tau^k_2, p^k_2)$ with the context tree models $(\tau^k_3, p^k_3)$ and $(\tau^k_4, p^k_4)$ might indicate that augmenting the number of contexts increases the learning difficulty. Accordingly, the context tree models $(\tau^k_3, p^k_3)$ and $(\tau^k_4, p^k_4)$ display much lower learning rates than those of the context tree models $(\tau^k_1, p^k_1)$ and $(\tau^k_2, p^k_2)$. 

We recall that the entropies of models $(\tau^k_1, p^k_1)$ and $(\tau^k_2, p^k_2)$ are both larger than the entropies of models $(\tau^k_3, p^k_3)$ and $(\tau^k_4, p^k_4)$. Taken together, these results clearly indicate that the entropy alone does not give an accurate indication of the difficulty of the learning task. 


Employing a statistical model selection procedure we were able to identify from the participants' responses both the set of contexts and the associated transition probabilities governing their predictions for the four different context tree models. We found that, for a fixed context tree, increasing the number of non deterministic transitions so as to increase the entropy of the chain deteriorated the participants' performance. Besides, disrupting the periodic structure of the stochastic sequence led to a performance worsening.

A performance worsening was also observed when the number of contexts used by the kicker increased. This is compatible with the findings of Wang et al. \cite{wang_learning_2017}, which investigated the dynamics of structure learning by tracking human responses to temporal sequences that change in structure without the participant's knowledge. In that study, the participants were asked to predict the upcoming item following a probabilistic sequence of symbols, ranging from independent transitions to context-based statistics. The participants succeeded in extracting the behaviorally relevant context length and transition probabilities corresponding to the structure of the presented sequence. Besides, increasing the context length worsened the performance.  

In another line of evidence, Kahn et al. \cite{Kahn2018} demonstrated that the graph architecture underlying probabilistic motor sequences fundamentally constrains learning. In their experiment, participants performed a probabilistic motor sequence task in which the order of button presses was determined by the traversal of different graphs (i.e., modular, lattice and random). Reaction times were shown to augment from modular to lattice and random graphs, indicating that probabilistic sequences learning is influenced by the graph topology. For instance, Kahn et al. \cite{Kahn2018} found that the inclusion of additional edges (with the same set of nodes) leads to an increase in the reaction time. This is equivalent to the transformation done in the graph representation of $(\tau^k_1, p^k_1)$ to $(\tau^k_2, p^k_2)$ (see Figure \ref{fig:exp_design}D), where we found a decrease in the proportion of correct guesses. Also, in Kahn et al. \cite{Kahn2018} the graph modular structure was associated with lower reaction times as compared to lattice and random graph structures. Likewise, in the present study the graph representation of the context tree model $(\tau^k_3, p^k_3)$ displays a modular structure that relates to the periodicity present in the sequences it generates. This modular structure was disrupted in the context tree model $(\tau^k_4, p^k_4)$ (see Figure \ref{fig:exp_design}D). As in Kahn et al. \cite{Kahn2018} the breaking of this modular structure could explain the worse performance found in $(\tau^k_4, p^k_4)$ as compared to $(\tau^k_3, p^k_3)$.   


In probability learning tasks, participants often start from a strategy of random guessing and may then slowly approach the maximizing behavior. Based on an extensive review \cite{MONTAG2021233}, Montag concluded that in simple probability learning tasks, participants normally tend to overshoot true matching behavior. In the present study for $(\tau^k_1, p^k_1)$ and $(\tau^k_2, p^k_2)$, most participants achieved a performance corresponding to the matching strategy from the first window of analysis on. However, for $(\tau^k_3, p^k_3)$ for most participants the proportion of correct predictions only attained the matching strategy from the fifth window of analysis on. Finally, for $(\tau^k_4, p^k_4)$, the proportion of correct predictions was below those of the matching scores in all time windows. This is an indicative that playing the game with $(\tau^k_1, p^k_1)$ and $(\tau^k_2, p^k_2)$ results in a simpler probability learning task than with $(\tau^k_3, p^k_3)$ and $(\tau^k_4, p^k_4)$.   

Comparing the performances obtained in $(\tau^k_1, p^k_1)$ and $(\tau^k_2, p^k_2)$ we observed that more participants went closer to the maximizing strategy in $(\tau^k_2, p^k_2)$ than in $(\tau^k_1, p^k_1)$. This is consistent with the findings of Wang et al. \cite{wang_learning_2017} showing that participants get closer to maximizing when the complexity of the task is increasing by changing the order of the stochastic chain from 0 to 1 and 2. In our case, this effect was observed after changing the transition probabilities associated to the contexts 01 and 21 that increased the entropy values associated to model $(\tau^k_2, p^k_2)$. Since the maximizing strategy was very rarely found in the case of the context tree models $(\tau^k_3, p^k_3)$ and $(\tau^k_4, p^k_4)$ the interplay between difficulty of the task and maximizing seems not linear. 


In the same vein our results suggest that entropy alone does not give an indication of the learning difficulty across context tree models. Context tree models $(\tau^k_3, p^k_3)$ and $(\tau^k_4, p^k_4)$, which are clearly more difficult to learn, have lower entropy than those of $(\tau^k_1, p^k_1)$ and $(\tau^k_2, p^k_2)$. A nonlinear relationship between entropy and complexity of Markov chains has already been pointed out \cite{Li1991}. Besides, alternative measures of complexity for stochastic chains and their relationship with entropy have been proposed \cite{Grassberger:1986, Bialek:2001}.        
   

In conclusion, our mathematical model assumes that the goalkeeper considers only the past choices of the kicker. This assumption is statistically supported by the data, once the goalkeeper has learned the model. Conversely, our data show that the goalkeeper's decision is influenced by its own past choices when he/she is still learning the context tree model structure. To the best of our knowledge, this is a new result that express an important feature of the goalkeeper's learning strategy.   

\section*{Methods}

\subsection*{Participants}	

A total of 122 healthy volunteers (60 female, mean age 31.95, standard deviation 9.41, right handedness 89.34\%, all of them with at least the secondary school level and only 6.56\% with no game familiarity) were recruited for the experiment. The volunteers signed an informed consent term after the objective of the study was explained to them. This experimental protocol was approved by the local ethics committee (Plataforma Brasil process number CAAE 58047016.6.1001.5261, statement approval number 1.846.941).

\subsection*{Experimental protocol}

The experiment was conducted remotely and consisted of two steps. In the first step, participants received a link containing an electronic form asking about gender, age, educational level, handedness and familiarity with electronic games, to characterize the sample of participants. After completing this form, the participants received a second e-mail containing the link to access the game. 

The game started with an instruction screen with the following statement: "You are a goalkeeper and you must predict before each kick in which region of the goal the ball will be kicked. The penalty taker may kick the ball toward the left, the center or the right of the goal. Your task is to defend the maximum of penalties. Attention, goalkeeper: the penalty taker is not influenced by your choices. During the experiment use the left, down and right arrows on the keyboard to choose the left, center and right side of the goal, respectively. During the game you will have two rest breaks that will be informed.
Have fun and improve your performance. The more penalties you defend, the better you will be ranked."

Four different context tree models were chosen to generate the sequences of kicker's choices (Figure \ref{fig:exp_design}B). Each participant was assigned to one of the four context tree models. As a result, the four groups consisted of 31, 30, 30 and 31 participants, respectively. Furthermore, within a context tree model group, each participant was exposed to a putative different sequence of kicker's choices.  

For each participant and each context tree model, a sample was constituted by collecting an ordered sequence of 1000 pairs, indicating the successive directions chosen by the penalty taker and the corresponding guesses of the goalkeeper. 

\subsection*{Structure of the sequence of stimuli}

The sequences of directions chosen by the kicker were generated by using a context tree model. A context tree model is characterized by a context tree representing the set of minima suffixes of the different sequences of past required to generate the next symbol. Associated to each context there is a transition probability indicating the probability with which the next symbol is chosen, given the context. For more details on context tree models we refer the reader to Rissanen (1983)\cite{Rissanen:83}, Buhlmann et al. (1999) \cite{Buhlmann-Wyner:99} and Galves et al. (2008)\cite{Galves-Loch:08}. 

In our experimental protocol, the four context tree models used to generate the sequences of the kicker's choices will be denoted by $(\tau^k_1, p^k_1)$, $(\tau^k_2, p^k_2)$, $(\tau^k_3, p^k_3)$ and $(\tau^k_4, p^k_4)$. In the above notation, the upper index $k$ stands for \textit{kicker} (Figure \ref{fig:exp_design}B).

The four context tree models are organized in pairs. In each pair the two models have the same context tree but with different associated families of transition probabilities (Figure \ref{fig:exp_design}B).

For the first pair ($(\tau^k_1, p^k_1)$ and $(\tau^k_2, p^k_2)$), the changes were done in the transition probabilities associated to contexts 01 and 21. In $(\tau^k_1, p^k_1)$, the transitions associated to these contexts are deterministic. In opposition, in $(\tau^k_2, p^k_2)$ we have two possible outcomes: with a small probability ($0.25$) we return to the same symbol which occurred before symbol 1, and with a high probability ($0.75$) we change to a new symbol. These changes increase the entropy from 0.65 for model $(\tau^k_1, p^k_1)$ to 0.81 for model $(\tau^k_2, p^k_2)$.   

In the second pair ($(\tau^k_3, p^k_3)$ and $(\tau^k_4, p^k_4)$), sequences generated by $(\tau^k_3, p^k_3)$ can be described as a concatenation of strings 211 in which the symbol 1 can be replaced by the symbol 0 with small probability (0.25) in an i.i.d way. Conversely, the sequences generated by $(\tau^k_4, p^k_4)$ can not be obtained by replacing symbols in a concatenation of deterministic strings in an i.i.d way (Figure \ref{fig:exp_design}C). This was achieved by interchanging the transition probabilities associated to contexts 01 and 21 and by changing the most probable outcome in the transition probability associated to context 2. These changes hardly changed the entropy (0.54 for model $(\tau^k_3, p^k_3)$ and 0.56 for model $(\tau^k_4, p^k_4)$).

\subsection*{Statistical Analysis}

In our experiment, the set of participants $V$ is divided in four groups according to the kicker context tree model $(\tau^k_i,p^k_i)$, $i=1,..,4$ that each participant is exposed to. For each participant $v = v(i) \in V$, $X_1^{v}, X_2^{v},...,X_{n}^{v}$ denotes the sequence of kicker choices and $Y_1^{v}, Y_2^{v},...,Y_n^{v}$ the participant responses recorded during the exposure to $(X^{v})_n$. 

In order to assess the temporal evolution of the performance, most of the statistical analysis were done using an sliding window approach. In this approach a window is formed over 250 trials, and this window slides over the data (150 trials) to capture different portions of it. Therefore we end up with six time windows. We denote by $(X_1^{v,j}, Y_1^{v,j}),...,(X_m^{v,j}, Y_m^{v,j})$ the data corresponding to the kicker choices and goalkeeper predictions for the participant $v \in V$ in the time window $j$, $j = \{1,...,6\}$.

\subsubsection*{Proportion of correct predictions}

The proportion of correct predictions (PCP) is defined as the proportion of times that the goalkeeper prediction matches the choice of the kicker in a defined set of trials $T$. Formally, 
\begin{equation}
	PCP_v = \frac{1}{|T|}\sum_{t=1}^{|T|} 1_{\{Y_t^{v} = X_t^{v}\}},
\end{equation}
where $|T|$ refers to the cardinal of the set $T$. 

The normalized proportion of correct predictions for a participant $v = v(i)$ computed in the set of trials $T$ is defined as the proportion of correct predictions on $T$ divided by the optimal theoretical score corresponding to the maximizing strategy for the kicker context tree model $i$.

When both, the proportion of correct predictions and the normalize proportion of correct predictions are computed across time windows, $|T|$ refers to the trials encompasses in the window.  

The curve of cumulative proportion of correct predictions corresponding to a goalkeeper $v$ is obtained by computing on each trial $t$ ($t \in \{1,...,1000\}$) the proportion of correct predictions considering the set $T = \{1,...,t\}$
\begin{equation}
	CPCP_v(t) = \frac{1}{t}\sum_{m=1}^{t} 1_{\left\{Y_m^{v} = X_m^{v}\right\}}.
\end{equation}

\subsubsection*{ANOVA}

Let $z_{ijv}$ be the normalize proportion of correct predictions of the participant $v$ playing the kicker context tree model  $(\tau_i^k, p_i^k)$ in the time window $j$ after a logit transformation. 

An univariate linear regression model was fitted to each participant normalized proportions of correct guesses $z_{ijv}$ as a function of time window $j$, i.e., $z_{ijv} = \gamma_{v}*j + b_v$. Participants with negative slope $\gamma_{v}$ were excluded from the statistical analysis. 

The mixed anova model fitted is,
\begin{equation}
	z_{ijv} = \mu + \alpha_j + \beta_i + (\alpha\beta)_{ij} + s_{(v(i))} + e_{ijv},
\end{equation}
where $\mu$ is the overall mean, $\alpha_j, j=1,...,6$ is the effect of the $j$th-level of the within subject factor (time window), $\beta_i, i=1,...,4$ is the effect of the $i$th-level of the between subject factor (kicker context tree model), $(\alpha\beta)_{ij}$ is the $ij$ interaction effect, $s_{(v(i))}$ is the random effect of the $v$th participant and $e_{ijv}$ is a random noise.

\subsubsection*{Statistical model selection procedure}

\begin{figure}[p]
	\centering
	\includegraphics[width = 0.9\textwidth]{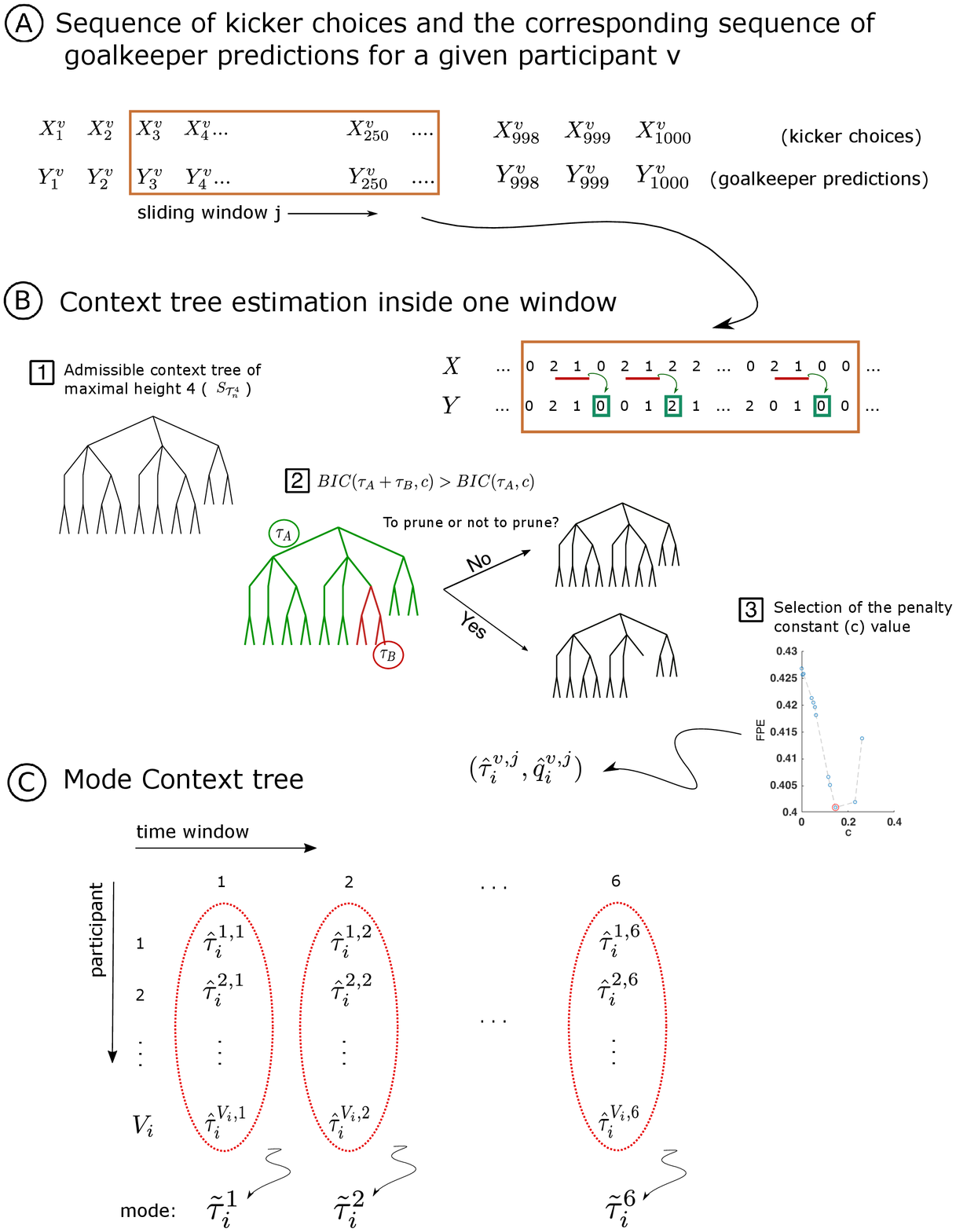}
	\caption{A) For each context tree and each participant, a sample consisted in an ordered sequence of 1000 pairs of events, each pair corresponding to the successive directions chosen by the kicker and the corresponding guesses of the goalkeeper. B) At each step, we use the string of past directions chosen by the kicker and the successive prediction made by the goalkeeper to estimate a transition probability. B1, B2) To retrieve the context tree used by the goalkeeper, we prune the tree of candidate contexts. Starting from the leaves, we prune the tree branches using the BIC criterion. B3) The penalty constant in the BIC is chosen so as to minimize the proportion of prediction errors \cite{Buhlmann-Wyner:99}. C) For each time window, the mode context tree was estimated from the retrieved set of context trees.}
	\label{fig:modelselection}
\end{figure}

For each participant $v \in V$ and each time window $j \in \{1,...,6\}$ the data $(X_1^{v,j}, Y_1^{v,j}),...,(X_m^{v,j}, Y_m^{v,j})$ is used to estimate the context tree $\hat{\tau}^{v,j}$ and the family of distributions $\hat{q}^{v,j}$ governing the predictions of the goalkeeper (see Figure \ref{fig:modelselection}A,B). The statistical procedure employed for this is summarized in the following algorithm.

\begin{algorithm}[H]
	\KwIn{The alphabet $A$, a sample $(x_1,y_1),...,(x_n,y_n)$  with $x_k,y_k \in A$ for $1 \leq k \leq n$, the maximum tree height $L$ (a positive integer), the penalization constant $c$ (a positive real number).}
	\KwOut{A context tree $\hat{\tau}$ and a family of distributions $\hat{q}$ indexed by the elements of $\hat{\tau}$.}   
	
	Compute the admissible context tree of maximal height $L$, $\mathcal{T}_n^L$ and its set of nodes $S_{\mathcal{T}_n^L}$
	
	\ForEach{$w \in S_{\mathcal{T}_n^L}$}{
		\tcp{compute recursively, beginning by the leaves, on each node}
		\tcp{the real value}
		\BlankLine
		$V_{w,n} = \left\{ \begin{array}{ll} n^{-c\cdot df(w)} L_w(y^n_1|x^n_1) & \textrm{, if $w$ is a leaf} \\
			\max\Big\lbrace n^{-c\cdot df(w)} L_w(y^n_1|x^n_1)\, , \, \prod_{b\in A}V_{bw,n} \Big\rbrace & \textrm{, otherwise}\end{array}  \right.$
		
		\BlankLine
		\tcp{the indicator}
		$\mathcal{X}_{w,n} = \left\{ \begin{array}{ll} 0 & \textrm{, if $w$ is a leaf} \\ \mathds{1}_{\Big\lbrace \prod_{b\in A}V_{bw,n} >  n^{-c\cdot df(w) } L_w(y^n_1|x^n_1) \Big\rbrace} & \textrm{, otherwise}. \end{array}
		\right.$
	}
	
	Select from $S_{\mathcal{T}_n^L}$ the nodes that have indicator equal 0 and all the nodes in the path from it to the root have indicator equal 1, that is $\hat{\tau} = \{w \in S_{\mathcal{T}_n^L}: \mathcal{X}_{w,n}=0 \mbox{ and } \mathcal{X}_{u,n}=1 \mbox{ for all  } u\prec w\}.$
	
	Compute $\hat{q}_w = \{\hat{q}(a|w), a \in A\}$, for each $w \in \hat{\tau}$.

	\Return{$\hat{\tau}, \hat{q} = \{\hat{q}_w, w \in \hat{\tau}\}$}
	
	\caption{Model selection algorithm for Sequence of random objects driven by context tree models based on BIC}
	\label{alg:context_tree_estimation}
\end{algorithm}

\vspace{1cm}

Some quantities used in the algorithm are defined as follows:
\[
L_{u}(Y_1^n\mid X^n_1) =\prod_{a \in A} \hat q(a|u)^{N_n^{XY}(u,a)},  
\]
\[
\hat{q}(a|u) = \frac{N_n^{XY}(u,a)}{N_n^X(u)} = \frac{N_n^{XY}(u,a)}{\sum_{a'\in A} N_n^{XY}(u, a')} .
\]
\[
N_n^{XY}(u,a) = \sum_{t = l(u)}^{n-1}\mathds{1}_{ \{ X_{t-l(u)+1}^t = u; Y_{t+1} = a \} }.
\]
\[
N_n^X(u) = \sum_{t = l(u)}^{n-1}\mathds{1}_{\{X_{t - l(u)+1}^t = u\}}.
\]
\[
df(w)= \sum_{a\in A} 1_{\{N_n^{XY}(w,a)\geq 1\}} - 1
\]

For each kicker's context tree model $(\tau_i^k, p_i^k)$, we have on each time window $j$ the set of context trees $ \mathcal{C}_{i,j} = \{\hat{\tau}^{v,j}, v \in V_i\}$, where $V_i$ is the subset of participants exposed to the kicker's context tree model $(\tau_i^k, p_i^k)$. We summarize this set of trees by computing the mode context tree $\tilde{\tau}_i^j$ (see Figure \ref{fig:modelselection}C). 

\FloatBarrier

\subsubsection*{Likelihood-ratio statistical tests for independence}  


A question of interest is to known whether the goalkeeper prediction $Y_n^v$ at trial $n$ is influenced only by the past kicker shootings $(X_1^{n-1})^v = (X_1^v,..., X_{n-1}^v)$ or by both, the past kicker shootings $(X_1^{n-1})^v$ and its own past plays $(Y_1^{n-1})^v = (Y_1^v,..., Y_{n-1}^v)$. To address this question, we consider the following testing problem,   
\begin{eqnarray*}
	H_0 &:& Y_n^v \mbox{ is conditionally independent of } (Y_{n-k}^{n-1})^v \mbox{ given } (X_{n-k}^{n-1})^v \qquad \mbox{ vs } \qquad \\
	H_1 &:& Y_n^v \mbox{ depends on } (X_{n-k}^{n-1})^v \mbox{ and } (Y_{n-k}^{n-1})^v,
\end{eqnarray*}
where $k$ refers to the length of the past considered in the sequences $X$ and $Y$.

Note that under the more general case, this model can be parameterized by a vector of transition probabilities $\theta = \{p(a|w_x,w_y), a \in A, w_x, w_y \in |A|^k\}$ taking values in a $(|A|-1)|A|^{2k}$-dimensional space $\Theta$. The parameters in the case of independence (i.e., under the null hypothesis) are in a subset $\Theta_0$ of the parameter space $\Theta$ in which the following restriction holds $\theta = \{p(a|w_x,w_y): p(a|w_x,w_y) = p(a|w_x) \quad \forall w_x, w_y \in |A|^k, a \in A \}$.

A common way to test nested hypothesis is by using the likelihood ratio test. Given the observed data $Y_1^n$, the likelihood ratio test statistics is computed by 
\begin{equation}
	R(Y_1^n) = \frac{\sup \{L(Y_1^n;\theta): \theta \in \Theta_0\}}{\sup \{L(Y_1^n;\theta): \theta \in \Theta\}},
\end{equation}
where $L(Y_1^n;\theta)$ is the likelihood of the sample $Y_1^n$ for the model specified by $\theta$. This statistics computes a ratio of the maximized likelihood of the sample under each of the hypothesis. It is known that, under the null hypothesis, as the sample size $n$ approaches infinity, the distribution of $-2\log(R_n)$ converges to a $\chi^2$ distribution with degree of freedom equal to the difference in dimensionality of $\Theta$ and $\Theta_0$ \cite{bibid}. The decision whether or not to reject the null hypothesis is then taken by comparing $-2\log(R(Y_1^n))$ to the $\chi^2$ value corresponding to a desired statistical significance $\alpha$.

In this case, the statistic is define by 
\begin{eqnarray}
	-2\log(R(Y_1^n)) &=& -2\left[\sum_{a \in A} \sum_{w_x \in |A|^{k'}} N_n(w_x,a) \log(\hat{q}(a|w_x)) - \right.\nonumber\\
	&&\left.\sum_{a \in A}\sum_{w_x \in |A|^{k'}}\sum_{w_y \in |A|^{k}} N_n(w_x,w_y,a)\log(\hat{q}(a|w_x,w_y))\right]
\end{eqnarray}
where $\hat{q}(a|w_x) = N_n(w_x,a)/	N_n(w_x)$ and $\hat{q}(a|w_x,w_y) = N_n(w_x,w_y,a)/N_n(w_x,w_y)$ are the maximum likelihood estimates of the parameters of the models restricted to $\Theta_0$ and $\Theta$ respectively; and   
$N_n(w_x,a) = \sum_{i=k'+1}^{n}\mathds{1}{\left\{X_{i-k'}^{i-1} = w_x; Y_i = a\right\}}$, $N_n(w_x,w_y,a) = \sum_{i=k'+1}^{n}\mathds{1}{\left\{X_{i-k'}^{i-1} = w_x; Y_{i-k}^{i-1} = w_y; Y_i = a\right\}}$, $N_n(w_x) = \sum_{a \in A}N_n(w_x,a)$, $N_n(w_x,w_y) = \sum_{a \in A}N_n(w_x,w_y,a)$.  

\section*{Data Availability}

The data and the codes used in the analyses are available upon request. 

\bibliography{references}

\section*{Acknowledgments} 

This work is part of University of S\~ao Paulo project Mathematics, computation, language and the brain, FAPESP project Research, Innovation and Dissemination Center for Neuromathematics (grant 2013/07699-0), project Plasticity in the brain after a brachial plexus lesion (FAPERJ grant E26/010002474/2016) and Financiadora de Estudos e projetos FINEP (PROINFRA HOSPITALAR grant 18.569-8). A.G and C.D.V. were partially supported by CNPq fellowships (grants 314836/2021-7 and 309560/2017-9, respectively). C.D.V. was also partially supported by a FAPERJ fellowship (CNE 202.785/2018), and by the Rio de Janeiro Neuroinflammation Research Network (FAPERJ grant E26/010.002418/2019). N.H. was successively fully supported by FAPESP fellowships (grants 2016/22053-7 and 2021/01551-7). We thank Bruno M. de Castro for many discussions in the beginning of this work and Marilucia Otama for helping with data collection.
	
\section*{Author contributions statement}

A.G. and C.D.V conceived the theoretical framework and the experimental protocol. C.D.V. and N.H. collected the data. A.G conceived The Goalkeeper Game and M.D.G led the software development team. A.G, C.D.V and N.H. analyzed the data and wrote the article. All authors reviewed the manuscript.

\section*{Additional information}

\textbf{Competing interests}: The authors declare no conflict of interest. 
		
\end{document}